# Introduction to Engineering Materials

*Ana Arauzo*
Universidad de Zaragoza, Pedro Cerbuna 12, 50009 Zaragoza, Spain

**Abstract**
This lecture presents an overview of the basic concepts and fundamentals of Engineering Materials within the framework of accelerator applications. After a short introduction, main concepts relative to the structure of matter are reviewed, like crystalline structures, defects and dislocations, phase diagrams and transformations. The microscopic description is correlated with physical properties of materials, focusing in metallurgical aspects like deformation and strengthening. Main groups of materials are addressed and described, namely, metals and alloys, ceramics, polymers, composite materials, and advanced materials, where brush-strokes of tangible applications in particle accelerators and detectors are given. Deterioration aspects of materials are also presented, like corrosion in metals and degradation in plastics.

## 1    Introduction

Materials are an important aspect of engineering design and analysis. The performance, reliability, and functionality of existing structures and devices are inherently determined by the properties of the materials from which they are constructed. Moreover, the discovery and development of novel materials continuously expand the boundaries of technological innovation, enabling the creation of new devices and applications. Historically, major periods in human development have been defined by the dominant materials of the time, such as the Stone Age, Bronze Age, and Iron Age, underscoring the centrality of materials in societal advancement. Today, we are entering what may be termed the era of Atomaterials: a new class of nanomaterials engineered through precise manipulation at the atomic scale, offering unprecedented control over material properties and performance.

Materials used in accelerator applications face numerous technological challenges. Addressing these challenges requires the integrated efforts of materials science and materials engineering. Materials science investigates the relation between structure and properties of materials whereas materials engineering applies this knowledge to design or engineer the structure of a material to produce a predetermined set of desired properties.

Selecting the most suitable material for a given application is a complex and non-trivial task. There is an enormous range of engineering materials with more than 500000 commercially available materials, each possessing unique properties and characteristics that make them suitable for specific applications. In-service conditions, deterioration during operation and cost of material and fabrication are the key factors to consider. The more familiar an engineer or scientist is with the various characteristics and structure-property relationships, as well as processing techniques of materials, the more proficient and confident he or she will be to make judicious materials choices.

In the science and engineering of materials, four fundamental components are critically interrelated: processing, structure, properties, and performance. Beyond structure and properties, processing and performance play essential roles in determining a material's suitability for specific applications. The structure of a material is directly influenced by the processing methods employed, while the material's performance is governed by its intrinsic properties. These relationships form a sequential and interdependent framework, often conceptualized as a linear progression: **processing → structure → properties → performance.**

Material selection for a specific application requires careful consideration of multiple factors across various property domains. Key considerations include: i) **Mechanical properties**: attributes such as strength, stiffness, toughness, and ductility are fundamental when designing load-bearing or mechanically active components; ii) **Thermal properties**: the material's thermal conductivity, heat capacity, and resistance to thermal degradation are critical in high-temperature environments; iii) **Chemical properties**: resistance to corrosion, oxidation, and other chemical interactions is essential for materials operating in aggressive chemical or environmental conditions; iv) **Electrical properties:** characteristics such as electrical conductivity, resistivity, and dielectric strength are vital in electronic and electrical applications; v) **Economic factors**: cost, material availability, and ease of processing or manufacturing significantly influence the feasibility and overall cost-efficiency of the final product. Additional physical properties, such as **optical and magnetic behaviour**, may also be relevant depending on the application. Importantly, many of these properties are strongly influenced by the material's microstructure, which is, in turn, determined by its composition and processing history.

## 2    Classification of materials

Some of the most common materials used in engineering include metals, polymers, ceramics, composites, and semiconductors. The engineering materials classification of solid materials is based on atomic structure with three basic groups: Metals and alloys (e.g., steel, aluminium), Polymers (e.g., polyethylene, nylon) and Ceramics and glasses (e.g., alumina, silica). Additional groups include composite materials (e.g., fiberglass, fiber-reinforced polymers), semiconductors and biomaterials and natural materials (e.g., wood, leather).

**Metals and alloys** are solid materials, typically hard, shiny, malleable, fusible, and ductile, with excellent electrical and thermal conductivity. These are broadly classified into two categories, ferrous, which are primarily iron-based, and non-ferrous, with other metal as the main component. Ferrous metals include steels, such as plain carbon steels and alloy steels, which contain less than 2% carbon, and cast irons, which typically contain between 2% and 4% carbon. These materials are widely used in structural and mechanical components due to their strength and versatility. Non-ferrous materials typically used in accelerators include Cu, Ni, Nb and Al alloys. Al and Ti alloys are frequently employed in vacuum applications. Ti alloys are also used for cryogenic applications and Highly pure Nb is a superconductor used for radiofrequency cavities, where Nb purity is a key factor for its performance. Tungsten (W) and molybdenum (Mo) alloys are used in beam pipes and lead (Pb) alloys in targets.

**Ceramics** are defined as any inorganic non-metallic material. They are classified as metal oxides (e.g., Alumina, Beryllia, Zirconia, glass), nitrides and carbides. Among these, advanced ceramics such as Silicon Carbide (SiC), exhibiting high thermal conductivity, excellent thermal stability and resistance to thermal shock, are commonly used in particle accelerators to dissipate heat effectively from high power components and maintain the integrity and performance of accelerator components [1].

**Polymers** are organic materials composed of long molecular chains formed by the repetition of smaller structural units known as monomers. There are two main groups based on their structure, thermo-softening plastic or thermoplastic and thermosetting polymers. **Thermoplastics** soften when heated and can be reshaped multiple times without undergoing significant chemical change. Upon cooling, they solidify and retain their new shape. Common examples include fluorocarbons (e.g., Teflon), polyamides (e.g., nylon), polyethylene, and polyesters (e.g., PET). In contrast, **thermosetting polymers** undergo a chemical transformation during curing, forming a rigid three-dimensional cross-linked structure that cannot be remelted or reshaped. This group includes elastomers (e.g., natural rubber, neoprene, silicones) and fiber-forming polymers (e.g., Lycra). In accelerator environments, materials and components are frequently exposed to intense radiation, which can induce significant changes in the physical and mechanical properties of polymers, particularly elastomeric materials. These changes may compromise their structural integrity and functionality. As a result, radiation resistance testing is essential for evaluating the long-term performance and suitability of polymers used in such high-radiation conditions.

**Composite materials** are engineered by combining two or more distinct constituents to produce a material with properties superior and/or different to those of the individual components. Composites can be categorized based on their supra-structural dimensionality into laminar, fibre-reinforced, and particle reinforced types. In particle accelerator facilities, composite materials are widely utilized in many parts and components. Examples are the high-modulus unidirectional carbon fiber profiles in the high-precision structures that support the silicon tracker modules in CMS experiment at CERN or the carbon fiber prototype tube part of the "High-Luminosity LHC upgrade of the CMS detector [2].

## 3   Structure of matter: microscopic vs macroscopic properties

The term structure refers to the spatial arrangement and organization of a material's internal components. Material structure can be examined and characterized across multiple length scales, typically categorized into five hierarchical levels:

i) Sub-atomic structure: the arrangement and behaviour of electrons and atomic nuclei, occurring at scales smaller than 1 angstrom (Å).

ii) Atomic structure: organization of atoms and molecules, typically on the order of 1-10 Å.

iii) Nanostructure: ranging from 1-100 nanometres (1-100 $10^{-9}$m).

iv) Microstructure: Large group of atoms, visible under optical or electron microscopy, spanning 0.1-100 micrometres (0.1-100 $10^{-6}$m).

v) Macrostructure: large-scale structure observable with the naked eye or low-magnification techniques, typically larger than 0.1 millimetres (> 1 $10^{-4}$m).

At sub-atomic level the electron energy states are described by quantum numbers, which define discrete electron shells and subshells within an atom. The electron configuration of an atom corresponds to the manner in which these shells and subshells are filled with electrons in compliance with the Pauli exclusion principle. The periodic table of the elements is generated by arrangement of the various elements according to valence electron configuration. Electropositive elements form positive ions while electronegative elements acquire electrons becoming negative ions. According to these properties the periodic table can be organised in the elements forming metallic materials, ceramic materials and polymeric materials.

### 3.1   Atomic bonding

Some of the important properties of solid materials depend on geometrical atomic arrangements, and also the interactions that exist among constituent atoms or Molecules. Atomic bonding in solids may be considered in terms of attractive and repulsive forces and energies. The three types of primary bond in solids are **ionic**, **covalent**, and **metallic**, which are determined by the electron structure of the individual atoms.

In **ionic bonding**, valence electrons are transferred from one atom to another, resulting in the formation of positively and negatively charged ions. The electrostatic (Coulombic) forces between these oppositely charged ions produce a strong, non-directional bond. Ionic bonding typically exhibits high bond energy and confers materials with characteristic properties such as hardness, brittleness, and poor electrical and thermal conductivity. This type of bonding is predominant in **ceramic materials**.

**Covalent bonding** involves the directional sharing of valence electrons between adjacent atoms along the axis connecting their nuclei. This type of bonding is characteristic of elemental solids found on the right side of the periodic table, such as carbon (C), silicon (Si), germanium (Ge), and compound semiconductors like gallium arsenide (GaAs). The strength of covalent bonds can vary significantly, ranging from relatively weak bonds (bismuth melts at 270 °C) to strong bonds (diamond melts at 3550 °C). Covalent bonding is typical of **polymeric materials**.

With **metallic bonding**, the valence electrons form a 'sea of electrons' that is uniformly distributed around a lattice of positively charged metal ion cores, binding them together. Typical of **metals and their alloys**. It is non directional, can be weak (Hg melts at -39ºC) or strong (Tungsten melts at 3410 ºC). Materials exhibiting metallic bonding generally possess high ductility and excellent electrical and thermal conductivity, properties that arise from the mobility of the delocalized electron cloud.

**Secondary bonding** forces arise from interactions between atomic or molecular dipoles, either permanent or induced, and are significantly weaker than primary chemical bonds. An electric dipole forms when there is a spatial separation of positive and negative charges within an atom or molecule. The resulting secondary bonding is due to the Coulombic attraction between the positive end of one dipole and the negative end of a neighbouring dipole. Secondary bonds include **van der Waals** forces and **hydrogen bonds**, both of which play critical roles in determining the properties of many materials, especially polymers. Van der Waals interactions are universal, present even in inert gases and molecular solids, and are fundamental in layered materials such as graphene and other two-dimensional (2D) materials. Hydrogen bonding is a specific, relatively strong form of dipole-dipole interaction involving hydrogen atoms bonded to highly electronegative atoms, which leads to elevated melting and boiling points compared to similar molecular structures lacking hydrogen bonds. In molecular liquids and crystals, the intermolecular forces are predominantly weak secondary bonds, resulting in relatively low melting and boiling temperatures. However, many modern polymers, despite being molecular materials composed of very large molecules, exist as solids due to their substantial molecular weight and chain entanglement, which enhance their mechanical stability.

## 3.2 Crystalline structures

The next hierarchical level of material structure corresponds to the geometrical arrangement of atoms in the solid state, commonly described by the crystal structure. The properties of some materials are directly related to their crystal structures. Permanent magnetic and ferroelectric behaviours of some ceramic materials are explained by their crystal structures. Moreover, significant differences in properties can arise between crystalline and non-crystalline (amorphous) forms of materials with identical chemical compositions. For example, non-crystalline ceramics and polymers normally are optically transparent; the same materials in crystalline (or semicrystalline) form tend to be opaque or, at best, translucent.

A crystalline material is one in which the atoms are situated in a repeating or periodic array over large atomic distances (long-range order). Each atom is bonded to its nearest-neighbour atoms. The unit cell is the basic structural unit or building block. All metals, many ceramic materials, and certain polymers form crystalline structures under normal solidification conditions. X-ray diffractometry is used for crystal structure and interplanar spacing determinations. A beam of x-rays directed on a crystalline material may experience diffraction (constructive interference) as a result of its interaction with a series of parallel atomic planes according to Bragg's law. Interplanar spacing is a function of the Miller indices and lattice parameter(s) as well as the crystal structure.

In amorphous or non-crystalline materials, long-range atomic order is absent. They have interesting properties like high mechanic strength and corrosion resistance. Plastics, glass, rubber, metallic glass, polymers, gel, fused silica, pitch tar, thin layer lubricants, and wax are examples of amorphous solids. Although there are various theoretical models, neither glass transition nor low-temperature properties of glassy solids are well understood on the fundamental physics level. Glass is a metastable phase that will change into a crystalline one if enough energy is provided.

Melt spinning is a widely used technique for producing amorphous metals, known as 'metglass'. This process typically yields thin ribbons, approximately 50 µm in thickness, with lengths extending to several meters and widths ranging from a few millimetres to several centimetres. In essence, the metal retains more or less its liquid and thus amorphous structure because it is cooled down with extreme speed (around 1.000.000 °C/s) supressing atomic diffusion during solidification. 'Bulk metallic glasses' represent an active area of research in Materials Science and Engineering.

**Metallic Crystal Structures** are non-directional in nature and present dense atomic packings. Most of the common metals crystallize in one of three principal crystal structures: body-centered cubic (BCC), face-centered cubic (FCC) and hexagonal Close-Packed (HCP).

**Ceramic Crystal Structures** comprise two or more elements with bonding ionic-covalent resulting in more complex structures. Two characteristics of the component ions in crystalline ceramic materials influence the crystal structure: the magnitude of the electrical charge on each of the component ions, and the relative sizes of the cations and anions. For a structure to be stable, the surrounding anions must be in contact with the central cation, which imposes a critical minimum cation-to-anion radius ratio for each coordination number. Examples include AX-type (A cation, X anion) such as NaCl, NgO, MnS, LiF, FeO, Fluorite type (e.g., $CaF_2$), Perovskite type (e.g., $BaTiO_3$), Spinel and inverse Spinel type ($MgAl_2O_4$, $Fe_3O_4$).

Some metals, as well as nonmetals, may have more than one crystal structure, a phenomenon known as polymorphism. When found in elemental solids, the condition is often termed allotropy. The prevailing crystal structure depends on both the temperature and the external pressure. This group of materials does not really fall within any one of the traditional metal, ceramic, polymer classification schemes. For example, graphite, one of the polymorphic forms of carbon, is often classified as a ceramic. Carbon is a particularly versatile element, capable of forming a wide range of allotropic structures, each with distinct bonding and dimensionality. Diamond, with a 3D structure $sp^3$ bonding is a good conductor of heat because of the strong covalent bonding and low phonon scattering. However, diamond is a metastable phase at ambient conditions (diamonds do not last for ever!) degrading to graphite 3D structure $sp^2$ bonding. Despite this, the transition back to graphite is kinetically hindered by a high energy barrier. Graphene, a 2D structure of $sp^2$-hybridized carbon atoms, and its derivatives are used as thermal conductive carriers to improve the thermal conductivity of thermal energy storage materials. Other carbon allotropes include carbon nanotubes (1D) and fullerenes such as $C_{60}$ (0D).

As with metals and ceramics, the properties of polymers are intricately related to the structural elements of the material. **Polymers** are mainly form of single chains of hydrocarbons with strong covalent bonds in each molecule and weak hydrogen and van del Waals bonds between molecules. The weak bonding between the chains defines the low melting and boiling points of linear polymers and its flexibility. When linear polymer chains are crosslinked, that is, chemically joined by covalent bonds between adjacent chains, a three-dimensional network is formed. These crosslinked polymers exhibit enhanced mechanical strength, hardness and dimensional stability compared to their non-crosslinked counterparts. Advances in modern characterization techniques have enabled the determination of the molecular structures of this group of materials, and the development of numerous polymers, which are synthesized from small organic molecules. Many of our useful plastics, rubbers, and fiber materials are synthetic polymers. The degree of crystallinity may range from completely amorphous to almost entirely (up to about 95%) crystalline. Many semicrystalline polymers form spherulites; each spherulite consists of a collection of ribbonlike chain-folded lamellar crystallites that radiate outward from its center.

Single crystals, a crystalline solid, where the periodic and repeated arrangement of atoms is perfect or extends throughout the entire specimen, are ordinarily difficult to grow, because the environment must be carefully controlled. If the extremities of a single crystal are permitted to grow without any external constraint, the crystal will assume a regular geometric shape having flat faces, as with some of the gem stones; the shape is indicative of the crystal structure. That is why most crystalline solids are composed of a collection of many small crystals or grains, forming a polycrystalline material. In single crystals, many physical properties can be anisotropic, varying with crystallographic direction depending on the symmetry of the crystal structure. In contrast, for polycrystalline materials, the measured properties represent an average over all crystallographic directions, unless grains have a preferential crystallographic orientation (texture).

## 3.3 Defects and dislocations

The properties of many materials are profoundly influenced by the presence of imperfections. Consequently, it is important to have a knowledge about the types of imperfections that exist, and the roles they play in affecting the behaviour of materials. For example, the mechanical properties of pure metals experience significant alterations when alloyed.

All materials have in a less or more extend the presence of **impurities** or foreign atoms. A pure metal consisting of only one type of atom does not exist. Metals to a purity of 99.9999% have $10^{22}$ to $10^{23}$ impurities/m$^3$. Most familiar metals are alloys, in which impurity atoms have been added intentionally to impart specific characteristics to the material. For instance, sterling silver is a 92.5% silver - 7.5% copper alloy. In normal ambient environments, pure silver is highly corrosion resistant, but also very soft. Alloying with copper enhances mechanical strength without depreciating corrosion resistance. The presence of carbon in iron significantly increases its strength and hardness, resulting in the creation of steel. Adding chromium to steel makes it resistant to corrosion. Bronze made of 80% Cu and 20% Sn is the first alloy in history and named an entire age of human civilization. Bronze is a substitutional alloy with FCC structure. The different size of tin atoms changes the structure and gives bronze many of its properties. Larger Sn atoms restrict the movement of Cu, making the material harder. Increasing the tin percentage to 22% reduces the vibration dampening of the material which is used for fabricating sonorous bells.

Crystallographic defects or structural imperfections are pivotal to material properties. Depending on the type, concentration and behaviour of these imperfections, mechanical, thermal and electronic properties are affected. There are three main types of structural imperfections attending to its dimensional classification: **point defects**, **line defects** and **surface imperfections**. Point defects are localized disturbances confined to one or two atomic sites, like vacancies, interstitials, and substitutional defects. Line defects, or dislocations, significantly affect the mechanical properties of materials and come in two forms: edge dislocations and screw dislocations. Surface imperfections include grain boundaries, twin boundaries, and stacking faults, disrupting the ideal geometrical arrangement over a significant region of the crystal. Understanding, controlling, and manipulating imperfections are fundamental in technological applications.

The simplest of the point defects is a **vacancy**, or vacant lattice site, one normally occupied from which an atom is missing. All crystalline solids contain vacancies, and in fact, it is not possible to create such a material that is free of these defects. The necessity of the existence of vacancies is explained using principles of thermodynamics; in essence, the presence of vacancies increases the entropy (i.e., the randomness) of the crystal. Self-interstitial is an atom from the crystal that is crowded into an interstitial site, a small void space that under ordinary circumstances is not occupied. In metals, a self-interstitial introduces relatively large distortions in the surrounding lattice because the atom is substantially larger than the interstitial position in which it is situated. Consequently, the formation of this defect is not highly probable, and it exists in very small concentrations, which are significantly lower than for vacancies. Materials of all types are often heat treated to improve their properties. The phenomena that occur during a heat treatment almost always involve atomic diffusion.

All metals and alloys contain some **dislocations** that were introduced during solidification, during plastic deformation, and as a consequence of thermal stresses that result from rapid cooling. The number of dislocations, or dislocation density in a material, is expressed as the total dislocation length per unit volume, or, equivalently, the number of dislocations that intersect a unit area of a random section. The units of dislocation density are millimetres of dislocation per cubic millimeter or just per square millimeter. Dislocation densities as low as $10^3$ mm$^2$ are typically found in carefully solidified metal crystals. For heavily deformed metals, the density may run as high as $10^9$ to $10^{10}$ mm$^2$. Heat treating a deformed metal specimen can diminish the density to $10^5$ to $10^6$ mm$^2$. By way of contrast, a typical dislocation density for ceramic materials is between $10^2$ and $10^4$ mm$^2$; also, for silicon single crystals used in integrated circuits the value normally lies between 0.1 and 1 mm$^2$.

Macroscopic plastic deformation in metals results from the movement of large numbers of dislocations. Therefore, a metal's ability to deform plastically depends on how easily dislocations can move. Since both hardness and strength (yield and tensile) are related to the ease of plastic deformation, restricting dislocation motion increases mechanical strength; that is, greater force is required to initiate deformation. In contrast, when dislocations move freely, the metal becomes softer and weaker. On the atomic level, plastic deformation involves breaking bonds with neighbouring atoms and forming new ones. In metals, this occurs primarily through a process called slip, which involves the motion of dislocations. Virtually all strengthening techniques rely on this simple principle: restricting or hindering dislocation motion renders a material harder and stronger.

For crystalline ceramics, plastic deformation also occurs by the motion of dislocations, but mobility is very restricted, therefore hardness and brittleness of these materials is partially due to the difficulty of dislocation motion. The main cause is predominant ionic bonding is restricting the slip by electrostatic repulsion. For ceramics in which the bonding is highly covalent, slip is also difficult: covalent bonds are relatively strong. They have limited numbers of slip systems and complex dislocation structures.

In polycrystalline materials, **surface imperfections** appear as two-dimensional interfacial defects known as grain boundaries. The grain structure plays a key role in determining the material's properties. For instance, materials with smaller grains are typically harder and stronger than those with larger grains because the increased grain boundary area more effectively hinders dislocation motion [3, 4].

## 3.4 Phase diagrams and transformations

Mechanical and other properties of many materials depend on their microstructures, which are often produced as a result of phase transformations. One reason why a knowledge and understanding of phase diagrams is important to engineers relates to the design and control of heat-treating procedures as some properties of materials are functions of their microstructures, and, consequently, of their thermal histories. Even though most phase diagrams represent stable (or equilibrium) states and microstructures, they are, nevertheless useful in understanding the development and preservation of nonequilibrium structures and their attendant properties; it is often the case that these properties are more desirable than those associated with the equilibrium state. For example, the tensile strength of an iron-carbon alloy of eutectoid composition (0.76wt% C) can be varied between approximately 700 MPa (100,000 psi) and 2000 MPa (300,000 psi) depending on the heat treatment employed.

Of all binary alloy systems, the one that is possibly the most important is that for iron and carbon. Both steels and cast irons, primary structural materials in every technologically advanced culture, are essentially iron-carbon alloys. Carbon steels account for 90% of total steel production. The microstructure that develops depends on both the carbon content and heat treatment. The molecular structure of pure iron at normal temperature is referred to as ferrite. This microstructure will also be found in steel with very low carbon content. A BCC crystal structure is the ferrite's distinguishing feature. Austenite is a microstructure generated when iron-based alloys are heated over 815 ºC but below 982 ºC. If the correct alloy, such as nickel, is present in the steel, the material will retain its microstructure even after cooling. Austenite is distinguished by its FCC crystal structure. The molecules in austenite are more densely packed than those in ferrite. When carbon steel is heated to austenite temperatures and subsequently cooled without any alloy present to maintain the austenite shape, the microstructure reverts to ferrite. However, if the carbon level exceeds 0.006%, the excess carbon atoms bond with iron to create iron carbide (Fe$_3$C), also known as cementite. Cementite does not form on its own since a portion of the material is ferrite. Pearlite is a laminated material composed of alternating layers of ferrite and cementite. It happens when steel is progressively cooled, generating a eutectic combination with alternating layers produced by a simultaneous crystallization of the two molten materials. Martensite has a tetragonal crystalline structure that is body-centered. This microcrystalline form is achieved by rapidly cooling steel, which traps carbon atoms inside the iron lattice. Ferritic, martensitic stainless steels are usually classified as 'magnetic', austenitic stainless steels are 'non-magnetic'. Cold working of some steels can induce a phase transformation from non-magnetic austenite (FCC) to magnetic martensite (BCT).

Cold working is commonly used as a strengthening method for metals. Since hardness and strength (both yield and tensile) are closely tied to the ease of plastic deformation, limiting dislocation mobility enhances mechanical strength. Strengthening techniques generally focus on restricting dislocation motion through various mechanisms. For single-phase metals, the main methods include grain size reduction, solid-solution alloying and strain hardening by cold working. Strain hardening phenomenon is explained on the basis of dislocation–dislocation strain field interactions. The dislocation density in a metal increases with deformation or cold work, due to dislocation multiplication or the formation of new dislocations. Consequently, the average distance of separation between dislocations decreases. On the average, dislocation-dislocation strain interactions are repulsive. The net result is that the motion of a dislocation is hindered by the presence of other dislocations. Alloys are stronger than pure metals because impurity atoms that go into solid solution ordinarily impose lattice strains on the surrounding host atoms. Lattice strain field interactions between dislocations and these impurity atoms result, and, consequently, dislocation movement is restricted.

# 4   Deterioration

Deteriorative mechanisms are different for the three material types. In metals, there is actual material loss either by dissolution (corrosion) or by the formation of non-metallic film (oxidation). Ceramic materials are relatively resistant to deterioration, which usually occurs at elevated temperatures or in rather extreme environments; the process is frequently also called corrosion. For polymers, mechanisms and consequences differ from those for metals and ceramics, and the term degradation is most frequently used. Polymers may dissolve when exposed to a liquid solvent, or they may absorb the solvent and swell; also, electromagnetic radiation (primarily ultraviolet) and heat may cause alterations in their molecular structure.

Corrosion may be defined as the destruction of a **metal** or an alloy because of chemical or electrochemical reaction with its surrounding environment or medium. Corrosion process is the reverse process of producing/extracting metal from ores, because pure metals are in a metastable state and tend to revert to their original state of oxides, sulphides, etc. This occurs because metals are electropositive and tend to an oxidated state, as a consequence most metal alloys experience corrosion and are biodegradable. Among metals, aluminium exhibits particularly good corrosion resistance. This is due to the formation of a thin, protective oxide layer (alumina, Al$_2$O$_3$) when the metal is exposed to oxygen. This passive layer acts as a barrier to further corrosion and has the unique ability to self-repair if damaged, as long as oxygen is available. In addition to its corrosion resistance, aluminium is the most abundant metal in Earth's crust, making up more than 8% by weight.

Corrosion of **ceramic materials** generally involves simple chemical dissolution. Ceramic are compounds between metallic and non-metallic elements, and therefore may be thought of as having already been corroded. As a result, ceramics exhibit exceptional resistance to further corrosion in most environments. This inherent stability makes them ideal for use in high-temperature and highly corrosive conditions.

In the case of **polymers**, degradation is physiochemical involving physical and chemical phenomena. Two processes produce degradation, dissolution and decomposition. Dissolution is produced by small solute molecules fitting into and occupying positions among the polymer molecules. Decomposition involves covalent bond rupture as a result of heat energy, chemical reactions, and radiation, reducing mechanical integrity. Irreversible degradation results when the temperature of a molten thermoplastic polymer is raised to the point at which molecular vibrations become violent enough to break the primary covalent bonds. Not all consequences of radiation exposure are deleterious. Crosslinking may be induced by irradiation to improve the mechanical behaviour and degradation characteristics.

Sometimes the degradation behaviour of a material for some application is overlooked, with adverse consequences. In order to define the strength of an engineering material for a corrosion-based design it is essential to define the nature of the environments affecting the material over time. Factors to be considered include, the material chemical composition, microstructure, grain boundary and surface condition, the stress factor, environment and thermodynamics and kinetics of the processes.

Several strategies can be employed to reduce corrosion in metals. When possible, use corrosion-resistant alloys; if not, apply protective coatings such as metallic (e.g., cadmium, chromium, nickel, aluminium, zinc) or paint coatings, the most common approach. Avoid contact between dissimilar metals, as this can lead to galvanic corrosion, where the less corrosion-resistant metal corrodes more rapidly due to electrical coupling. To further reduce corrosion risk, minimize surface defects such as scratches, inclusions, oxides, and grain boundaries, which act as high-energy sites for corrosion initiation. Additionally, ensure that residual salts and chemical contaminants, often introduced during processing or surface preparation, are thoroughly removed, as they can accelerate degradation.

## 5 Applications in accelerators

There are more than 30,000 particle accelerators in operation around the world. Most of them use low energy beams ($\ll$ 1 MeV) and are used in industry. Less numerous are high energy accelerators for research where there are just over a hundred. Particle physics requires pushing the accelerated beams to the highest possible energies and to the highest possible intensities.

Environmental conditions for particle accelerators are very different from industrial environment. Materials may be subjected to cryogenic temperatures (magnets and components), ultra-high vacuum (down to $10^{-11}$ mbar) high magnetic fields, intense radiation, elevated temperatures (in beam intercepting components like collimators and targets) and high strain rates. Material properties are strongly temperature ($T$) and magnetic field ($H$) dependent, yet most commercial suppliers do not test materials under such extreme conditions. Virtually all physical properties, thermal, electrical, magnetic, and mechanical, vary with temperature. Additionally, many degradation and failure mechanisms are governed by time- and energy-dependent processes, further influencing long-term performance.

The next future envisages going beyond the frontiers of high-energy particle physics with the construction of new and more powerful accelerators, increasing demands on the materials used for the accelerating structures: intense radiofrequency fields, great radiation dose, higher magnetic field. In particular, superconducting magnets are required for many of the proposed facilities, in some cases with operational parameters well beyond current state-of-the-art [5]. Superconducting Nb-Ti magnets needed for colliders operate with fields up to 8.3 T (in the LHC). Dipole fields in excess of 16 T (up to 24 T) are needed, and solenoids, primarily for the muon collider, up to 50 T. A recently tested US MDP prototype $Nb_3Sn$ magnet has reached 14.5 T.

New challenges in particle accelerators demand application-specific material requirements and specifications. Advanced composites, amorphous and polycrystalline materials, additive manufacturing, and functional coatings offer promising solutions, but also introduce new complexities. In accelerator engineering, failure is not an option or at least the failure possibilities have to be minimal. Success relies on careful material selection, understanding of processing effects, system-level design, and ensuring components are integrated and used appropriately within their application.

## References

[1] https://www.wits.ac.za/news/latest-news/research-news/2021/2021-05/south-african-materials-technology-used-in-cerns-atlas-experiment.html.

[2] A. Jung and C. Pierce, "Enormous carbon fiber tube crafted at Purdue will support future hunt for discoveries at CERN," https://www.physics.purdue.edu/interactions/2022/research/jung_cms.php.

[3] C. Xu, H. He, Z. Xue, and L. Li, "A detailed investigation on the grain structure evolution of AA7005 aluminum alloy during hot deformation," *Mater Charact*, vol. 171, p. 110801, Jan. 2021, doi: 10.1016/J.MATCHAR.2020.110801.

[4] R. G. Guan and D. Tie, "A review on grain refinement of aluminum alloys: Progresses, challenges and prospects," May 01, 2017, *Chinese Society for Metals*. doi: 10.1007/s40195-017-0565-8.

[5] S. Gourlay, T. Raubenheimer, and V. Shiltsev, "Challenges of Future Accelerators for Particle Physics Research," Jun. 27, 2022, *Frontiers Media SA*. doi: 10.3389/fphy.2022.920520.

## Bibliography

Callister, W. D. and Rethwisch D. G. (2009), Materials Science and Engineering: An Introduction, 8th Edition; John Wiley and Sons. ISBN-10: 0470419970; ISBN-13: 978-0470419977.

Callister, W. D. and Rethwisch D. G. (2011), Fundamentals of Materials Science and Engineering: An Integrated Approach, 4th Edition; Wiley. ISBN-10: 1118061608; ISBN-13: 978-1118061602.

Smith, W. and Hashemi J. (2009), Foundations of Materials Science and Engineering, 5th Edition; McGraw-Hill Science/Engineering/Math. ISBN-10: 1118061608; ISBN-13: 978-0073529240.

Shackelford, James F., Introduction to Materials Science for Engineers, 8th Edition; Pearson. ISBN-10: 0133826651; ISBN-13: 978-0133826654.

Shackelford, James F., Materials Science and Engineering Handbook, 3rd Edition; CRC Press. ISBN 0-8493-2696-6.

Yuli K. Godovsky, Thermophysical Properties of Polymers, Springer-Verlag 1992, DOI 10.1007/978-3-642-51670-2.

Online Resources:

https://ocw.mit.edu/courses/3-012-fundamentals-of-materials-science-fall-2005/

http://www.istl.org/02-spring/internet.html.